\documentclass[a4paper,12pt]{article}
\usepackage[pctex32]{graphics}

\newcommand{\be}{\begin{equation}}
\newcommand{\ee}{\end{equation}}
\newcommand{\ben}{\begin{eqnarray}\displaystyle}
\newcommand{\een}{\end{eqnarray}}

\begin{titlepage}

\begin{document}
{\baselineskip20pt


\vskip .6cm

\begin{center}
{\Large \bf Thermodynamics and phase transitions in the
Born-Infeld-anti-de Sitter black holes}

\end{center} }

\vskip .6cm
 \centerline{\large Yun Soo Myung$^{1,a}$,
 Yong-Wan Kim $^{1,b}$,
and Young-Jai Park$^{2,c}$}

\vskip .6cm

\begin{center}
{$^{1}$Institute of Basic Science and School of Computer Aided
Science,
\\Inje University, Gimhae 621-749, Korea \\}

{$^{2}$Department of Physics, Sogang University, Seoul 121-742, Korea}
\end{center}

\vspace{5mm}


\begin{abstract}
We investigate the Born-Infeld-anti-de Sitter black hole
(BIAdS) solutions in the four dimensions, which is a nonlinear
generalization of the Reissner-Norstr\"om-AdS black hole (RNAdS).
We analyze all thermodynamic quantities of the BIAdS in the
canonical ensembles, which are characterized by the charge $Q$,
the mass $M$, the nonlinear parameter $b$, comparing with those of
the RNAdS and Schwarzschild-AdS black hole. We find the forbidden
region of $0\le bQ <0.5$ for the presence of a charged BIAdS. We
also discuss the Hawking-Page phase transitions in the BIAdS black
holes. Here we obtain a new Hawking-Page phase transition for the
$bQ=0.5$ critical BIAdS.
\end{abstract}


\vskip 0.8cm

\noindent $^a$ysmyung@inje.ac.kr \\
\noindent $^b$ywkim65@gmail.com \\
\noindent $^c$yjpark@sogang.ac.kr

\noindent
\end{titlepage}

\setcounter{page}{2}

\section{Introduction}

Since Born and Infeld (BI) proposed a nonlinear electrodynamics
~\cite{BI}, BI action has received renewed attention in string
theory~\cite{FTL,Tse,Lei}. For various motivations, extending the
Reissner-Nordstr\"om  black hole (RNAdS) in Einstein-Maxwell
theory to the charged black hole solutions in Einstein-Born-Infeld
(EBI) theory has attracted much attention in recent years, for example,
see~\cite{GSP,Dem,Wil,Ras,Tt,GH,Bre,AFG,Dey,Fern,She}.
In particular, Einstein gravity in (2+1)
dimensions~\cite{Car,Man,FHR} has been intensively studied because
of the existence of black holes solutions in (2+1)-dimensional
anti-de Sitter (AdS) spacetimes~\cite{btz,btz1,mtz,CG}, which
possess certain features inherent to the (3+1)-dimensional black
holes. Recently, we have systematically obtained all thermodynamic
quantities of the EBI black holes in three dimensions by comparing
those of the Maxwell and BTZ black holes~\cite{mkp3}. These are
all between non-rotating uncharged black hole (NBTZ) and charged
black hole (CBTZ)~\cite{Ida}. This result in 3D EBI black holes (BIBTZ)
may provide new insights towards a better understanding of the (3+1)-dimensional
Born-Infeld-anti-de Sitter black holes (BIAdS), whose
thermodynamic analysis is regarded as a
nontrivial task to carry out completely ~\cite{Dey,Fern}.

On the other hand, since the pioneering work of Hawking-Page on
the phase transition between thermal AdS and AdS black hole in
four dimensions~\cite{HP}, the research of the black hole
thermodynamics has greatly improved additionally suggesting that
there may exist a different phase transition between small and
large black holes (HP1) in the RNAdS for a fixed charge
$Q<Q_c$~\cite{CEJM,DMMS1,Myu} and AdS Gauss-Bonnet black
holes~\cite{Cho,DMMS2,NO,CNO}. In contrast, in the conventional
Hawking-Page phase transition (HP2), one generally starts with
thermal radiation in AdS space appearing an unstable small black
hole  with negative heat capacity. Then, since the heat capacity
changes from negative infinity to positive infinity at the minimum
temperature, the large black hole with positive heat capacity
finally comes out as a stable object. There is a change of the
dominance at the critical temperature: from thermal radiation  to
black hole~\cite{HP}.

In this paper, we address these issues for the BIAdS, which is a
nonlinear generalization of the RNAdS. We completely analyze all
thermodynamic quantities of the BIAdS in the canonical ensemble,
which is characterized by charge $Q$, mass $M$, and the nonlinear
parameter $b$. Note that the work of the BIAdS in the grand
canonical ensemble was carried out in Ref. \cite{Fern}. Moreover,
we obtain a new type of the Hawking-Page phase transition (HP3) in
the BIAdS black holes, which has been still not observed in the
previous works~\cite{HP,CEJM,DMMS1,Myu,Cho,DMMS2,fk,cpw,MO}.

The organization of this work is as follows. In Sec. 2, we
carefully analyze the possible BIAdS black hole solutions, which
are classified by the charge $Q$ and the nonlinear parameter $b$.
In Sec. 3, we investigate all thermodynamic properties of the
allowed BIAdS as a nonlinear realization of the RNAdS. In Sec. 4,
we present a new phase transition in the BIAdS black holes. We
summarize and discuss our results in Sec. 5. In Appendix, we
derive the free energy using the Euclidean action approach.

\section{Structure of BIAdS black holes}

Now, let us consider a (3+1)-dimensional gravity coupled with
nonlinear electrodynamics known as the BIAdS
action~\cite{Dey,fk,cpw}
\begin{equation}\label{action}
S=\int d^4x \sqrt{-g}\left(\frac{R-2\Lambda}{16\pi G}+L(F)\right),
\end{equation}
where
\begin{equation}\label{actionF}
L(F)=\frac{b^2}{4\pi G}\left(1-\sqrt{1+\frac{2F}{b^2}}\right)
\end{equation}
with $F\equiv F^{\mu\nu}F_{\mu\nu}/4$.
Here, the constant $b$ is the Born-Infeld parameter, which is
related to the string tension $\alpha'$ as $ b = 1/(2 \pi
\alpha')$, and $\Lambda = - 3/l^2$ is the cosmological constant.
Hereafter we choose $G=1$ for simplicity.
Note that this Lagrangian reduces to the RNAdS one in the limit
$b^2 \rightarrow \infty$.

By solving the equations of motion for the gauge field $A_\mu$ and
the gravitational field $g_{\mu\nu}$,
the BIAdS black hole solutions~\cite{Dey,fk,cpw} can be written as
\begin{equation} \label{metric}
ds^2 = - f(r) dt^2 + f(r)^{-1} dr^2 + r^2  d \Omega^2.
\end{equation}
Here, the metric function $f(r)$ is given by
\begin{eqnarray}
f(r) &=& 1 - \frac{2M}{r} + \frac{r^2}{l^2} + \frac{2 b^2 r^2}{3} \left( 1 - \sqrt{ 1 + \frac{Q^2}{b^2r^4 }} \right)  \nonumber \\
&{}& +~\frac{ 4 Q^2}{ 3 r^2} \hspace{0.2cm}  {\cal F} \left(
\frac{1}{4}, \frac{1}{2}, \frac{5}{4}, -\frac{Q^2}{b^2 r^4 }
\right),
\end{eqnarray}
where ${\cal F}$ is a hypergeometric function \cite{abra},
and only non-zero component with the electric charge $Q$ is given by
$F^{01} = -E  = - Q/{\sqrt{ r^4 + {Q^2}/{b^2}}}$.
Hereafter we only consider $Q\ge 0$ and $b\ge0$ without any loss of generality.
At this stage we note two limiting cases as guided black holes to study
the BIAdS black holes.  In the limit $Q \rightarrow 0$,
$f(r)$ reduces to the Schwarzschild-anti de Sitter black hole (SAdS) case,
while in the limit $b \rightarrow \infty$ and $Q \neq 0$, $f(r)$ reduces to the RNAdS case.
However, the BIAdS should have the main feature of the
RNAdS: two horizons $r_\pm$ and extremal black hole because the BIAdS is a
nonlinear generalization of the RNAdS. This implies that the
SAdS with single horizon
does not belong to the category of the RNAdS and thus it should be
disconnected to the BIAdS.

On the other hand, the ADM mass $M$ defined by $f(r) = 0$ is given by
\begin{eqnarray}
\label{aas1} M(r_+,Q,b) &=& \frac{r_{+}}{2} + \frac{r_{+}^3}{2l^2}
+ \frac{ b^2 r_{+}^3}{3}
\left( 1 - \sqrt{ 1 + \frac{Q^2}{ b^2 r_+^4}} \right) \nonumber \\
&{}& +~\frac{ 2 Q^2}{ 3 r_+} \hspace{0.2cm}  {\cal F} \left(
\frac{1}{4}, \frac{1}{2}, \frac{5}{4}, -\frac{Q^2}{ b^2 r_+^4}
\right)
\end{eqnarray}
with the outer horizon $r=r_+$.
In order to describe the BIAdS with $Q \neq 0$ and $b \neq 0$,
which is the generalization of the RNAdS, we first have to know
the extremal BIAdS. In this case, both $f(r)$ and
$\frac{df(r)}{dr}$ have to be zero at the degenerate horizon. Such conditions
will lead to the equation
\begin{equation}
1 + \left( 2 b^2 + \frac{3}{l^2} \right) r_{e}^2 - 2 b^2 \sqrt{
r_{e}^4 + \frac{Q^2}{b^2}} =0.
\end{equation}
The solution is given by
\begin{equation}
r_{e}^2 = \frac{l^2}{6}\left(\frac{1 + \frac{3}{2b^2 l^2}}{1 +
\frac{4}{4b^2 l^2}}\right) \left[- 1 + \sqrt{ 1 + \frac{12\left(1
+ \frac{3}{4b^2 l^2}\right)}{b^2l^2\left(1 + \frac{3}{2b^2 l^2}
\right)^2} \left(b^2Q^2 -\frac{1}{4} \right)}~\right].
\end{equation}
Then, we require the condition $ b Q \ge 0.5$ in order to have a
real root for $r^2_{e}$. Note that $r^2_e=0$ for $ b Q = 0.5$,
while $r^2_e$ reduces to the RNAdS in the limit $b\rightarrow\infty$.
On the other hand, for $ 0 \le b Q < 0.5$, we could not find the extremal
BIAdS.  Hence, we call this  as the forbidden region for the
BIAdS. As a result, we propose that the meaningful parameter space
for the BIAdS is not the whole, but
\begin{equation}
0.5 \leq bQ \leq \infty.
\end{equation}
We call the lower bound as the critical BIAdS and the upper bound of
$b \to \infty$ as the RNAdS. This  lower bound could be
understood from the fact that the presence of $b$ leads to the
screening effects on the electric field $E$, and thus makes the
role of charge less important. Hence, for $0\le bQ<0.5$, any
charged black hole does not exist.

\begin{figure}[t!]
   \centering
   \includegraphics{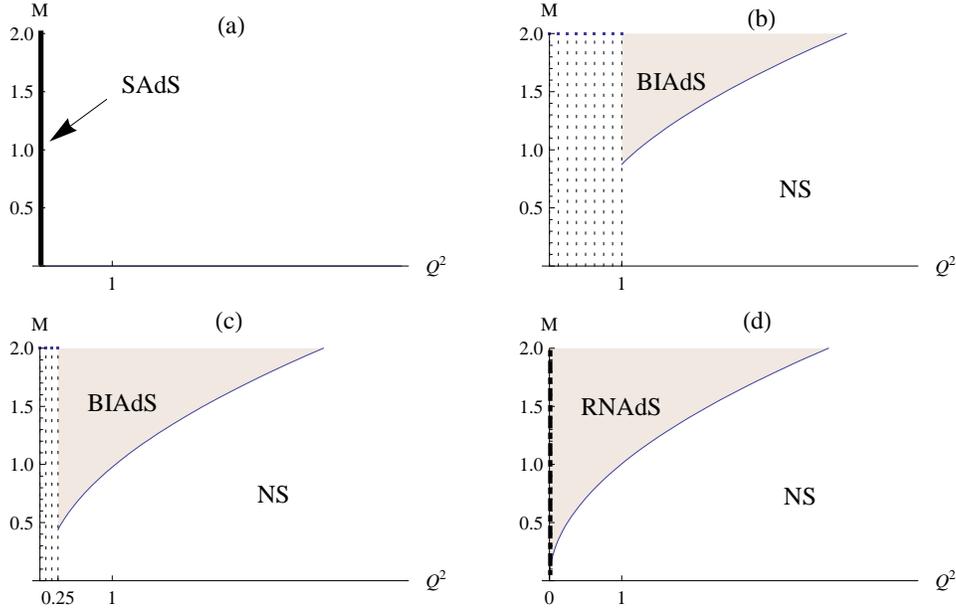}
\caption{The mass function graphs $M$ vs $Q^2$ with $l=10$: (a)
the SAdS case ($Q=0$), two BIAdS cases with
(b) $b=0.5$ and (c) $b=1.0$, and (d) the RNAdS case, respectively.
The naked singularity regions (NS) are
white areas below solid curves of the extremal BIAdS, while the
forbidden regions (FR) are dotted areas between positive $M$-axis ($Q=0$)
and $bQ=\frac{1}{2}$.}
\label{fig1}
\end{figure}

Next, for the case of $ b Q \geq 0.5$, the mass of the extremal black
hole is given by
\begin{equation}
\label{extremass}
M_{e} \equiv M(r_e,Q,b),
\end{equation}
which reduces to the well-known RNAdS case in the limit of $b
\rightarrow \infty$. As is shown in Fig. 1, when $b$ increases
($b=0.5 \to 1 \to \infty$), the dotted forbidden regions (FR) are
getting narrow and narrow  while naked singularity regions (NS)
are increasing wide and wide.

Let us discuss the behavior of mass as a function of horizon
radius $r=r_\pm$. First, we consider the case of $bQ>0.5$. If $ M
> M_{e}$, then there will be two horizons: the inner $r_-$ and
outer $r_+$. For $ M = M_{e}$ there will be a degenerate horizon
of $r_-=r_+=r_e$. If $M < M_{e}$, there are no horizons as shown
in Fig. 2, and it will yield a naked singularity. Second, in the
case of $bQ=0.5$, $r_e=0$ is the extremal point and for $M>M_e$,
one has one horizon $r_+$ only. Finally, there exists a dotted
forbidden region described by $ 0 \le bQ <0.5$ including the SAdS,
which can not be allowed for the BIAdS having the limiting case of
the RNAdS. This can be easily checked by expanding $M(r_+,Q,b)$ in
series near $r_+=0$ as
\begin{equation}
M(r_+,Q,b)\approx M_0 -\Big(bQ-\frac{1}{2}\Big)r_+ +
\left(\frac{b^2}{3}+\frac{1}{2l^2}\right)r_+^3 + {\cal O}[r_+]^5
\end{equation}
with
\begin{equation}
M_0=\frac{\Gamma[\frac{1}{4}]^2}{6\sqrt{\pi}}\sqrt{bQ^3}.
\end{equation}
This shows clearly that there exists a non-zero mass $M_0$
depending on $b,Q$ even at $r_+=0$. In fact, $M_0$ is hard to be
interpreted as the mass of a black hole. Therefore, for the case
of $ 0 \le  b Q < 0.5$ (the positive slop of linear term), we
could not obtain the BIAdS black hole solution having  the
extremal configuration. However, for the case of $bQ > 0.5$, the
slop of the linear term is negative, explaining how the extremal
configurations could be being developed. For the case of $bQ=0.5$,
the linear term disappears and its case is marginal.
\begin{figure}[t!]
   \centering
   \includegraphics{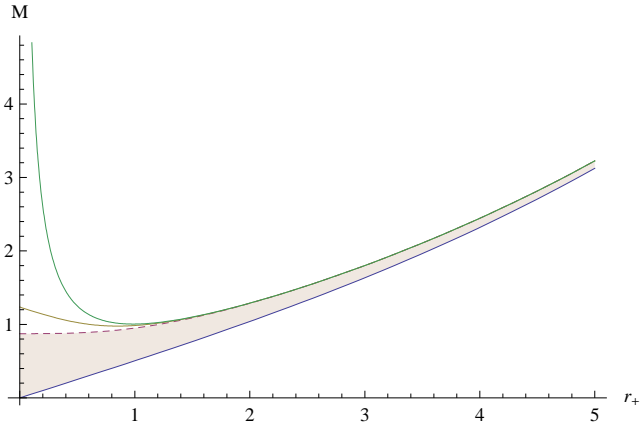}
\caption{The mass function graphs $M$ vs $r_+$ with $l=10$ from
bottom to top: the SAdS case ($Q=0$), two BIAdS cases
($Q=1$) with $b=0.5$ and $b=1.0$, and the RNAdS case,
respectively. The dotted line stands for $b$=0.5.
The shaded region represents the forbidden region of BIAdS: $0\le bQ
<0.5$. } \label{fig2}
\end{figure}

\section{Thermodynamics of  BIAdS black holes}

Now, we are ready to analyze the BIAdS black hole, which is really
a nonlinear generalization of the RNAdS  in the canonical
ensembles.
\begin{figure}[t!]
   \centering
   \includegraphics{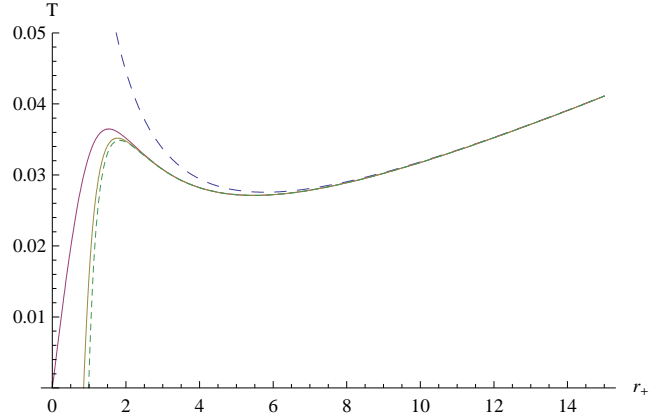}
\caption{Temperature graphs $T$ vs $r_+$ with $l=10$: the dashed
line is for the SAdS case ($Q=0$), two BIAdS cases ($Q=1$) with
$b=0.5$, $b=1$, and the dotted line for the RNAdS case,
respectively.} \label{fig3}
\end{figure}
The Hawking temperature  defined by $T_H=f'(r_+)/4\pi$ takes the
form
\begin{equation}
\label{aas2} T_H(r_+,Q,b) = \frac{1}{4\pi} \left( \frac{1}{r_+} +
\frac{3r_+}{l^2} + 2 b^2 r_+ \left[1 - \sqrt{1 + \frac{Q^2}{b^2
r_+^4}}~\right]  \right).
\end{equation}
We note that in the limit $Q \rightarrow 0$,  $T_H$ reduces to the
SAdS case, while in the limit of $b \rightarrow \infty $ and $Q
\neq 0$, $T_H$ reduces to the well-known RNAdS case. Then, using
the Eqs. (\ref{aas1}) and (\ref{aas2}), the heat capacity
$C(r_+,Q,b)=(dM/dT_{H})_Q$ for a fixed-charge is obtained to be
\begin{equation}\label{aac1}
C = \frac{2 \pi r^2_+ \sqrt{1 + \frac{Q^2}{b^2 r^4_+}} \left[ 3
r^4_+ + l^2 \left(r^2_+ + 2 b^2 \left(1 - \sqrt{1 + \frac{Q^2}{b^2
r^4_+}}\right) r^4_+\right) \right]} { 3 \sqrt{1 + \frac{Q^2}{b^2
r^4_+}} r^4_+ + l^2 \left[- \sqrt{1 + \frac{Q^2}{b^2 r^4_+}}
r^2_{+} +2 Q^2  - 2 b^2 \left( 1 - \sqrt{1 + \frac{Q^2}{b^2
r^4_+}} \right) r^4_+ \right]}. \end{equation}
The proper on-shell free energy defined by $F(r_+,Q,b)= M-M_e -T_{H}S_{BH}$
with the Bekenstein-Hawking entropy $S_{BH} = \pi r^2_+ $ is given by
\begin{eqnarray}
\label{aaf1} F &=& \frac{r_{+}}{4} - \frac{r_{+}^3}{4l^2} - \frac{
b^2 r_{+}^3}{6} \left( 1 - \sqrt{ 1
+ \frac{Q^2}{ b^2 r_+^4}} \right)  \\
&+& \frac{ 2 Q^2}{ 3 r_+} \hspace{0.2cm} {\cal F} \left(
\frac{1}{4}, \frac{1}{2}, \frac{5}{4}, -\frac{Q^2}{ b^2 r_+^4}
\right) - M_e. \nonumber
\end{eqnarray}
Here, we choose the extremal black hole as background because we
are working with the fixed-charge $Q$ ensemble~\cite{CEJM}. In
order to prove the correctness of free energy,  $F$ will be
derived from the Euclidean action approach in the Appendix. Also,
the free energy (\ref{aaf1}) may be  obtained   using  the
2-dimensional dilaton gravity  through the dimensional reduction
~\cite{GKV,GMc,MKPm,mkp3}.

Note that  in the limit $Q \rightarrow 0$, the heat capacity $C$ and
free energy $F$ reduce to the SAdS case as
\begin{eqnarray}\label{aaSc}
C^{SAdS}(r_+)&=& 2\pi r_+^2 \Big[\frac{3r_+^4+l^2 r_+^2}{3r_+^4 - l^2 r_+^2}\Big], \\
\label{aaSf} F^{SAdS}(r_+)&=& \frac{r_{+}}{4}\Big(1 -
\frac{r_{+}^2}{l^2}\Big),
\end{eqnarray}
where $C^{SAdS}$ blows up at $r_0=l/\sqrt{3}$ and $F^{SAdS}=0$ for
$r_+=0, l$. In the limit of $b^2 \rightarrow \infty$,  these
reduce to the well-known RNAdS case as follows:
\begin{eqnarray}\label{aac2}
C^{RNAdS}(r_+,Q)&=& 2\pi r_+^2 \Big[\frac{3r_+^4+l^2(r_+^2-Q^2)}{3r_+^4+l^2(-r_+^2+3Q^2)}\Big], \\
\label{aaf2} F^{RNAdS}(r_+,Q)&=& \frac{r_{+}}{4}\Big(1 -
\frac{r_{+}^2}{l^2} + \frac{ 3 Q^2}{  r_+^2}\Big) - M^{RNAdS}_e.
\end{eqnarray}
\begin{figure}[t!]
   \centering
   \includegraphics{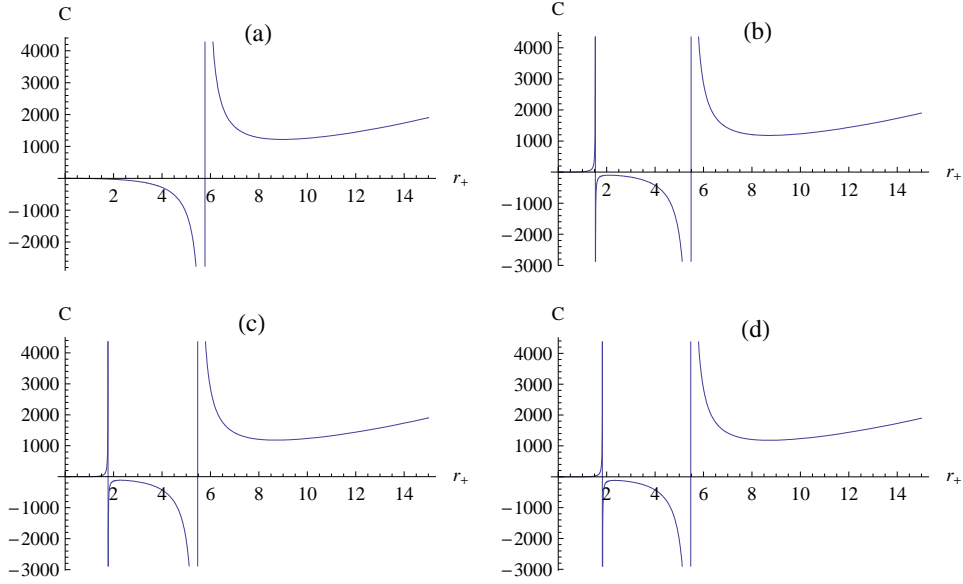}
\caption{Specific heat graphs $C$ vs $r_+$ with $l=10$: (a) the
SAdS case ($Q=0$), two BIAdS cases ($Q=1$) with (b)
$b=0.5$, (c) $b=1.0$, and (d) the RNAdS case,
respectively.} \label{fig4}
\end{figure}
The global features of the Hawking temperature depending on the
parameter $b$ are shown in Fig. 3. It seems to be a combination of
the RN and Schwarzschild black holes in the AdS space for the
cases of $b\geq0.5$. Here we observe the local minimum $T_H=T_0$
at $r_+=r_{0}$ (the feature of the SAdS black hole), and the
maximum value $T_H=T_m$ at $r_+=r_{m}$ and $T_H=0$ at $r_+=r_e$
(the feature of the RN black hole). In addition, the position
$r_+=r_e$ having the extremal temperature $T_H=0$ is being shifted
to zero when $b$ decreases from $b\to\infty$ (RNAdS) to $b=1$
(BIAdS), and then to $b=0.5$ (BIAdS). This implies that the
nonlinear effects of the BI action make the extremal position
close to zero, similar to the BIBTZ in three
dimensions~\cite{mkp3}.

The graphs of the heat capacity depending on the parameter $b$ are
shown in Fig. 4. For the SAdS case of $Q=0$, we have the typical form of heat
capacity, showing the change from $-\infty$ to $\infty$ at the
minimum temperature point of  $ r_0=l/\sqrt{3}$. Hence, this case
has simply two phases of negative and positive heat capacities. On
the other hand, for the BIAdS case of $bQ \ge 0.5$, we have the
RNAdS-type heat capacities with two discontinuities at $r_+=r_m,
r_0$ corresponding to the solution of $(dT_H/dr_+)_{Q}=0$ and zero
at $r_+=r_e$ related to the solution of $T_H=0$. In particular,
for $r_e<r_+<r_m$, the black hole is locally stable because of
$C>0$ while for $r_m<r_+<r_0$, it is locally unstable ($C<0$). For
$r_+>r_0$, the black hole becomes stable because of $C>0$. As a
result, we observe that $C=0$ at $r_+=r_e$, and it blows up at
$r_+=r_m,~r_0$. Based on the local stability, we introduce the
stable small black hole (SBH$_+$) with $C > 0 $ being in the region of
$r_e<r_+<r_m$, the unstable intermediate black hole (IBH$_-$) with $C<0$ in the
region of $r_m<r_+<r_0$, and the stable large black hole (LBH$_+$) with $C>0$
in the region of $r_+>r_0$.

\section{Phase transitions in BIAdS black holes}
As is well-known, the free energy plays a crucial role to test the
phase transition. The global features of the on-shell free energy
depend on the parameter $b$. We find that the nonlinear
generalization with $bQ \ge 0.5$ of the Einstein-Maxwell theory
does not change the thermodynamic stability in the canonical
ensemble. That is, the cases of $b=0.5,~1.0$ could not make the
free energy lower than the RNAdS with $b \to \infty$ in contrast
to the case of the grand canonical ensemble~\cite{Fern}, which
states that the BIAdS is thermodynamically preferred (stable) over
the RNAdS. This is mainly because we choose the extremal black
hole as the ground state in the canonical ensemble~\cite{CEJM}.
The only change is shifting  of extremal points toward the origin
as $b$ decreases.

In order to investigate what kinds of the Hawking-Page phase
transition are possible to occur in the BIAdS, we have to study
the off-shell process of the growth of a black hole. For this
purpose, we introduce the off-shell free energy as follows
\begin{equation} \label{offf}
F_{off}(r_+,Q,b,T) = M(r_+,Q,b) - M_e(Q,b) - T S_{BH}(r_+)
\end{equation}
with the  external temperature  $T$ of heat reservoir.
 We describe the phase transitions by considering four cases: $Q=0$ SAdS,  $
 bQ=0.5$ critical BIAdS, $bQ=1$ BIAdS, $b\to\infty$ RNAdS.
 We newly classify the Hawking-Page phase transitions according to
the previous works~\cite{DMMS1,DMMS2}.
\begin{figure}[t!]
   \centering
   \includegraphics{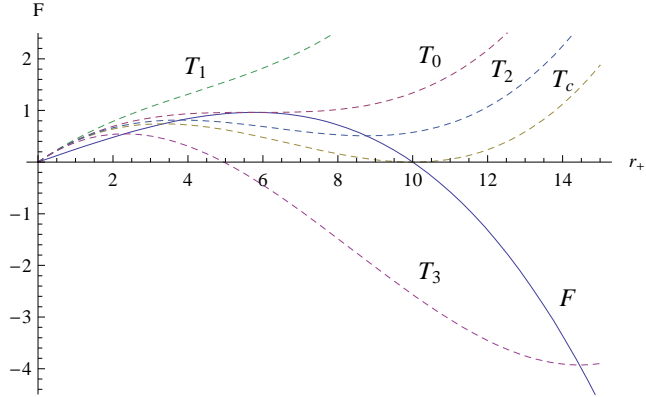}
\caption{Phase transition for the SAdS: minimum temperature
$T_{0}=0.02757$ at $r_{0}=5.774$, critical temperature
$T_c=0.03183$ for HP2 at $r_c=10$. In addition, three external
temperatures $T_1(=0.02)$, $T_2(=0.03)$, and $T_3(=0.04)$ are
introduced for the phase transition. The solid curve represents
the on-shell free energy $F^{SAdS}(r_+)$, while five dashed curves
denote the off-shell free energy $F^{SAdS}_{off}(r_+,T)$ with five
temperatures. } \label{fig.5}
\end{figure}

\subsection{$Q=0$ SAdS black hole (HP2)}
 A Schwarzschild-AdS black hole (SAdS) is globally stable only when $C>0$ and
$F<0$.  We observe that the free energy is   maximum at
$r_0=l/\sqrt{3}$ and zero at $r_c=l$ as shown in Fig. 5. For
$r_+>r_c$, one finds negative free energy. The
 temperatures correspond to  minimum $T_0=T_H^{SAdS}(r_0)$
and $T_c=T_H^{SAdS}(r_c)$, respectively.  Using the off-shell free
energy
\begin{equation}
F_{off}^{SAdS}=M^{SAdS}-T S_{BH}
\end{equation}
we find the phase transition.  For $T=T_3$, the process of phase
transition is shown in Fig. 5. In this case, one generally starts
with thermal radiation ($r_+=0$) in AdS space appearing an
unstable small black hole (SBH$_-$)  at $r_+=r_u$ (solution to
$F_{on}^{SAdS}(r_u)=F^{SAdS}_{off}(r_u,T_3))$ with negative heat
capacity. Here SBH$_-$ denotes the unstable small black hole with
$C<0$ and $F>0$. Then, since the heat capacity changes from
negative infinity to positive infinity at $r_+=r_0$ (see Fig.
4-a), the large black hole (LBH$_+$) with $C>0$ finally comes out
as a stable object at $r_+=r_s$ (solution to
$F^{SAdS}(r_s)=F^{SAdS}_{off}(r_s,T_3))$. Here LBH$_+$ denotes the
stable large black hole with $C>0$ and $F<0$. Actually, there is a
change of the dominance at the critical temperature $T_c$: from
thermal radiation  to black hole~\cite{HP}. This is called the
conventional Hawking-Page phase transition (HP2): AdS $\to$
SBH$_-$ $\to$ LBH$_+$. For the $T=T_1, T_2(<T_c)$ cases, the free
energy of AdS space $F^{SAdS}(0)=0$ is the lowest one, while for
the $T=T_3(>T_c)$ case, the lowest one is the negative free energy
$F^{SAdS}(r_s)$ of the large black hole.

\subsection{$bQ= 0.5$ critical BIAdS black hole (HP3)}
\begin{figure}[t!]
   \centering
   \includegraphics{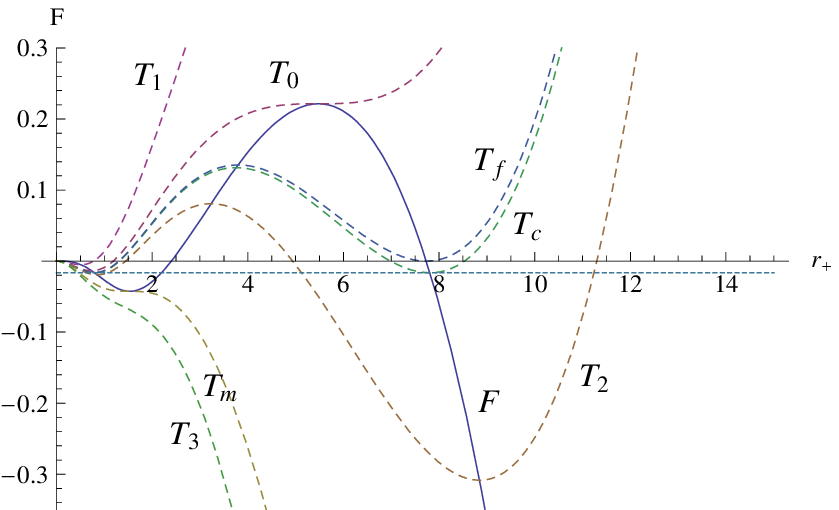}
\caption{Phase Transition for the $b=0.5$ critical BIAdS with
$Q=1$: $T_m=0.03647$ at $r_m=1.531$, local minimum $T_0=0.02712$
at $r_0=5.478$, $T_f=0.02858$ at $r_f=7.734$, $T_c=0.02867$ at
$r_s=0.796$ and $r_l=7.810$, $T_1(=0.02)$, $T_2(=0.03)$,
$T_3(=0.04)$. The solid curve represents the on-shell free energy
$F(r_+,Q=1)$, while seven dashed curves denote the off-shell free
energy $F_{off}(r_+,Q=1,T)$ with seven temperatures.}
\label{fig.6}
\end{figure}
Here we discuss the critical BIAdS case with $bQ=0.5$ as a new
phase transition. From Fig. 6, the local minimum $F=F_{min}$ is at
$r_+=r_{m}$,  the maximum value $F=F_{max}$ is  at $r_+=r_{0}$,
and $F=0$ at $r_+=r_e(=0)$ and $r_f$. The negative free energy
decreases in the range of $r_e<r_+<r_m$, and then it increases in
the region of $r_m<r_+<r_0$. For $r_+>r_0$, the positive free
energy again decreases to zero at $r_+=r_f$, and finally becomes
negative for $r_+ > r_f$. On the other hand, from Fig. 4-b, we
find the change of the heat capacity: $C=0$ at $r_e=0$; $C>0$ for
$0<r_+<r_m$; $C<0$ for $r_m<r_+<r_0$; $C>0$ for $r_+>r_0$. Here,
we define the extremal black  hole (EBH$_0$) at $r_+=r_e$ with
$T_H=C=F=0$.

From  Fig. 6, we can propose a new type of the Hawking-Page phase
transition (HP3). In this case, the critical temperature $T=T_c$
is derived from the condition of
$F_{off}(r_s,1,T_c)=F_{off}(r_l,1,T_c)$, which means that the free
energy of SBH$_+$ at  $r_+=r_s$ is equal to free energy of LBH$_+$
at $r_+=r_l$.  Here we distinguish SBH$_+$ with SBH$_-$: the
former is a stable small black hole with $C>0,F<0$, while the
latter is unstable small black hole with $C<0,F>0$. Then we expect
that a HP3 occurs between EBH$_0$ and LBH$_+$ through SBH$_-$ and
IBH$_-$ for $T_m<T<T_c$. Here IBH$_-$ represents the unstable
intermediate black hole with $C<0,F>0$. All of SBH$_+$, IBH$_-$,
and LBH$_+$ are the saddle points as the solutions to $F=F_{off}$.
Particularly, we choose $T_2=0.03$ for a transition temperature of
the HP3. This transition consists of three processes: EBH$_0$
$\to$ SBH$_+$ $\to$ IBH$_-$ $\to$ LBH$_+$. We have global
stabilities for SBH$_+$ and LBH$_+$ because of their free energies
are negative and heat capacities are positive. However, the
IBH$_-$ remains unstable. It may play a role of a mediator for HP3
as a SBH$_-$ in HP2 because of $C<0$ and $F>0$. Here we observe
that the dominance of system is changed from SBH$_+$ to LBH$_+$ at
$T=T_c$, called the critical temperature of HP3. For $T<T_c$, the
free energy of SBH$_+$ is the lowest one, while for $T>T_c$, the
lowest free energy is for the LBH$_+$.

It seems appropriate to comment on $T=T_f$, which may be considered
as the critical temperature. However, this shows the unwanted sequence of
$F_{off}(r_s)<F_{off}(r_e)=F_{off}(r_l)(=0)<F_{off}(r_i)$, which
implies that the transition from EBH$_0$ to LBH$_+$ unlikely
occurs because the free energy of SBH$_+$ is the lowest one. Hence,
we should choose $T=T_c$, which shows the desired sequence of
$F_{off}(r_s)=F_{off}(r_l)<F_{off}(r_e)(=0)<F_{off}(r_i)$. This case
implies that a new type of the Hawking-Page phase transition
from EBH$_0$ to LBH$_+$ is possible to occur.

\subsection{$bQ=1$ BIAdS black holes (HP1)}
\begin{figure}[t!]
   \centering
   \includegraphics{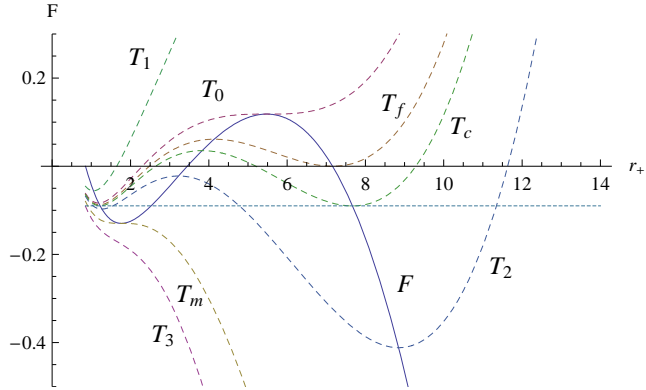}
\caption{Phase Transition for the $b=1$ BIAdS with $Q=1$:
$T_m=0.03518$ at $r_m=1.769$, local minimum $T_0=0.02712$ at
$r_0=5.478$, temperature $T_f=0.02799$  at $r_f=7.159$,
$T_c=0.0285$ at $r_s=1.224$ and $r_l=7.700$, $T_1(=0.02)$,
$T_2(=0.03)$, $T_3(=0.04)$. The solid curve represents the
on-shell free energy $F(r_+,Q=1)$, while seven dashed curves
denote the off-shell free energy $F^{off}(r_+,Q=1,T)$ with seven
temperatures.} \label{fig.7}
\end{figure}

Now, let us discuss the BIAdS case with $bQ=1$. From Fig.
\ref{fig.7}, the local minimum $F=F_{min}$ is at $r_+=r_{m}$ and
the maximum value $F=F_{max}$ is  at $r_+=r_{0}$. The free energy
is negative for $r_e<r_+<r_m$ and it increases in the region of
$r_m<r_+<r_0$. For $r_+>r_0$, it decreases to zero at $r_+=r_f$
and becomes negative for $r_+>r_f$. On the other hand, from Fig.
4-c, we find the change of heat capacity: $C=0$ at
$r_+=r_e\not=0$; $C>0$ for $0<r_+<r_m$; $C<0$ for $r_m<r_+<r_0$;
$C>0$ for $r_+>r_0$.

From  Fig. 7, we can propose another phase transition
(HP1)~\cite{DMMS1,DMMS2}. Here the off-shell free energies always
start with the nonzero values at $r_+=r_e$ because of
$S_{BH}(r_e)\not=0$. This contrasts to the critical BIAdS. In this
case, the critical temperature $T=T_c$ is derived from the
condition of $F_{off}(r_s,1,T_c)=F_{off}(r_l,1,T_c)$, which means
that the free energy of SBH$_+$ at  $r_+=r_s$ is equal to free
energy of LBH$_+$ at $r_+=r_l$. Then we expect that  a HP1 occurs
between SBH$_+$ and LBH$_+$ through IBH$_-$ for $T_m<T<T_c$.
Particularly, we choose $T_2=0.03$ for a transition temperature of
the HP1. This transition consists of two processes: SBH$_+$ $\to$
IBH$_-$ $\to$ LBH$_+$. We have global stabilities for SBH$_+$ and
LBH$_+$ because of their free energies are negative and heat
capacities are positive. However, the IBH$_-$ remains an unstable
mediator for HP1. Here we observe that the dominance of system is
changed from SBH$_+$ to LBH$_+$ at $T=T_c$, called the critical
temperature of HP1. For $T<T_c$, the free energy of SBH$_+$ is the
lowest one, while for $T>T_c$, the lowest free energy is for the
LBH$_+$.

\subsection{$b\to\infty$ BIAdS black hole (RNAdS:HP1)}
This case is similar to the $bQ=1$ BIAdS case. There is a phase
transition like HP1: SBH$_+$ $\to$ IBH$_-$ $\to$ LBH$_+$ because
of $r^{RNAdS}_e\not=0$~\cite{Myu}. This occurs for the case of
$Q<Q_c$ in the canonical ensemble of the RNAdS. For the case of $Q
\ge Q_c$, the transitions are not allowed in the RNAdS as the
limiting case of the BIAdS~\cite{CEJM}.

\section{Summary and discussions}

In summary, we have considered the four dimensional BIAdS, which
is the nonlinear generalization having the limiting case of the
RNAdS. We have carefully analyzed all possible BIAdS black hole
solutions, which depend on the charge $Q$ and the nonlinear
parameter $b$ with the reality condition $bQ \geq 0.5$. Then, we
have obtained all thermodynamic properties of the possible BIAdS
comparing with those of the Reissner-Norstr\"om-AdS and
Schwarzschild-AdS black holes.

Moreover, we have obtained possible Hawking-Page phase transitions
in the BIAdS black holes through the analysis of the off-shell
process of the growth of a black hole by introducing the off-shell
free energy. The first one is HP2 of  AdS $\to$ SBH$_-$ $\to$
LBH$_+$, which belongs to the conventional Hawking-Page phase
transition in the Schwarzschild-AdS black holes. For the critical
BIAdS, we have found a new phase transition (HP3) through the
process of EBH$_0$ $\to$ SBH$_+$ $\to$ IBH$_-$ $\to$ LBH$_+$. This
is similar to the phase transition in the non-rotating BTZ black
hole (NBTZ) of EBH$_0$ $\to$ NBTZ$_+$ in three
dimensions~\cite{Myungplb}. The difference is the number of
different phases: single phase for the NBTZ and three phases for
the case of the $bQ=0.5$ critical BIAdS. The phase transition for
the case of the $bQ>0.5$ BIAdS is HP1, which shows the transition
between SBH$_+$ and LBH$_+$ as SBH$_+$ $\to$ IBH$_-$ $\to$
LBH$_+$. For the RNAdS, which is the $b\to\infty$ BIAdS,
one has the similar phase transition like HP1.

At this stage, let us discuss whether the HP1 is really possible
to occur because this is the phase transition between two globally
stable objects (SBH$_+$ and LBH$_+$). It was reported that the HP1
is unlikely to occur in the RNAdS~\cite{Myu}. This is because the
maximum temperature of $T=T_m$ is present at the Davies' point of
$r_+=r_m$~\cite{Dav}. Whether the phase transition at $r_+=r_m$ is
possible or not is a long-standing issue to be
resolved~\cite{Pavon,JP,Rmkp}. The presence of maximum temperature
gives rises to a stable small black hole of SBH$_+$. If the
transition between SBH$_+$ and IBH$_-$ is not possible, two
transitions of HP3 and HP1 except HP2 are unlikely possible to
occur. In this work, we have discussed all phase transitions  by
assuming  that the phase transition between SBH$_+$ and IBH$_-$ is
likely to occur.

Now we ask whether the case of $bQ=0.5$ BIAdS is really a critical
point for the transition to  the Schwarzschild-AdS black holes. As
was pointed out previously, it is not the case because  the BI
action (BIAdS) is a nonlinear generalization of Maxwell theory
(RNAdS). The key point of the RNAdS is that it has an extremal
point of zero temperature when compared with the SAdS. Hence, we
insist that discussing the transition to the SAdS is irrelevant to
the BIAdS.

Finally, we discuss the implications of our results to the string
theory. According to the AdS/CFT correspondence, the Hawking-Page phase
transition of HP2 corresponds to the confining-deconfining phase
transition on the CFT side~\cite{Witt}. Because the BI action is
related to the string theory when choosing $b=1/2\pi \alpha'$, the
phase transitions of HP3 and HP1 could be applied to explore the
presumed transitions in the CFT.

\medskip
\section*{Acknowledgments}
Y. S. Myung  was supported by the Korea Research Foundation (KRF-2006-311-C00249)
funded by the Korea Government (MOEHRD). Y.-W. Kim was supported by the Korea
Research Foundation Grant funded by Korea Government (MOEHRD):
KRF-2007-359-C00007. Y.-J. Park was supported by
the Korea Science and Engineering Foundation (KOSEF) grant
funded by the Korea government (MOST) (R01-2007-000-20062-0).

\section*{Appendix: Free energy from the Euclidean action approach}

In the appendix, we derive  the free energy in Eq. (14)  from the
Euclidean action approach~\cite{CEJM,Fern}. In general, there are
two approaches: the grand canonical thermodynamic ensemble for the
fixed-potential and canonical thermodynamic ensemble for the
fixed-charge. For the AdS space,  the background consists of both
charged and uncharged quanta free to fluctuate in the presence of
fixed-potential $\Phi$. In the  case of grand canonical ensemble
at fixed-temperature and fixed-potential, the relevant
thermodynamic potential is the Gibbs free energy defined by
$W=M-T_HS_{BH}-\Phi Q$. The proper free energy for the
fixed-potential was derived in Ref. \cite{Fern}.

On the other hand, for the fixed-charge $Q$, the AdS space with a
fixed-charge is not a solution to the BIAdS equations. In this
case, in order to keep the fixed-charge, we use the extremal black
hole with the mass $M_e$ as the background and retain only
uncharged (neutral) quanta in the heat reservoir. Here, the
relevant quantity is the Helmholtz free energy defined by
$F=E-T_HS_{BH}$ with $E=M-M_e$.

In this appendix, we derive the desired free energy for the fixed-charge
from the Euclidean action approach.
In order to compute the free energy for the case of a fixed-charge
$Q$ canonical ensemble, let us consider the Euclidean action
\begin{equation}\label{actionT}
I_E=I_b+I_{GHY}+I_s-I_c-I_e
\end{equation}
Here $I_b$ is the Euclidean version (bulk term) of $S_0$ in
Eq. (\ref{action}), while the remaining terms are all boundary
terms. These  boundary terms except $I_e$ fall into two classes:
boundary terms $I_{GHY}$ and $I_s$,
which are required to get the correct boundary value problem,
belong to type I, while counter-term $I_c$, which is required to get the
correct variational principle, belongs to type II.

Among boundary terms, $I_{GHY}$ is the Gibbons-Hawking-York term
\begin{equation}
I_{GHY}=\int_{\partial M}d^3x
   \sqrt{h}K,
\end{equation}
where $K$ is the trace of extrinsic curvature. When the space is
asymptotically AdS, $I_{GHY}$ gives a vanishing
contribution~\cite{CEJM}.

Upon variation of the gauge field, the boundary terms from the
gauge field will vanish if one keeps the potential
$A_t(\infty)=\Phi$ fixed. However, if we wish to keep the charge
fixed, we must add a boundary surface term
\begin{eqnarray}
\label{surfT} I_s&=&-\frac{1}{4\pi G}\int_{\partial M}d^3x
   \sqrt{h}\left(\frac{F^{\mu\nu}}{\sqrt{1+2F/b^2}}\right)
   n_\mu A_\nu,
\end{eqnarray}
where $h_{ij}$ is the induced metric on the boundary at $r=r_B$
and $n_{\mu}$ is a radial unit vector pointing outwards.

The counter term $I_c=I_{AdS}=-(\Lambda/8\pi G)\int_0^{r_B}
d^4x\sqrt{g}$ is the Euclidean action of the pure AdS space, which
is necessary to regularize the divergence on the boundary at
infinity. This is the standard AdS counter term.

The last term $I_e$ is necessary to have the action for the
fixed-charge ensemble, using the extremal black hole as the
background. In order to compute the action, we evaluate
$I_b+I_{GHY}+I_s-I_c$ for a black hole of mass $M>M_e$ and then
subtract the contribution from the extremal background. This is
because the ground state of the BIAdS for the fixed-charge is not
the AdS space but the extremal black hole defined by the thermal
condition of  $T_H=0, C=0, F=0$. The extremal black hole  is
considered as a small, stable remnant against evaporating process
via the Hawking radiation. Hence, we may consider $I_c(I_e)$ the
IR (UV) counter-terms to have a proper, thermodynamic system.

Now, we are ready to obtain the proper Euclidean action (\ref{action}).
First, let us calculate the bulk term. Considering the equation of
motion for $g_{\mu\nu}$
\begin{equation}
R_{\mu\nu} - g_{\mu\nu}\Lambda  =
\frac{1}{2}g_{\mu\nu}\left(-{\cal L}(F)
    +\frac{\partial {\cal L}}{\partial g^{\rho\sigma}}g^{\rho\sigma}\right)
    -\frac{\partial {\cal L}}{\partial g^{\mu\nu}},
\end{equation}
the bulk term  can be rewritten as
\begin{equation}
\label{bulkS} I_b=-\frac{1}{16\pi G}\int_{M}d^4x\sqrt{g}
   \left(2\Lambda-{\cal L}(F)-\frac{2F^2} {\sqrt{1+F^2}}
   \right).\\
\end{equation}
After substituting  the electric field $F_{r\tau}$ into $I_b$ with
\begin{equation}
F_{r\tau}=i F_{rt}=\frac{-iQ}{\sqrt{r^4+\frac{Q^2}{b^2}}},
\end{equation}
it could be integrated to give  the form
\begin{equation}
I_b=\frac{\omega\beta}{16\pi G}
    \left[\frac{2}{l^2}(r_B^3-r_+^3)+\frac{4b^2}{3}(r_B^3-r_+^3)-4b
    \int_{r_{+}}^{r_B}\sqrt{r^4+Q^2}\right].
\end{equation}
Here $\omega$ is the volume of the two-sphere and $\beta$ is the
inverse temperature.

Second, the surface term $I_s$ is computed to have
\begin{equation}
I_s=\frac{\omega\beta}{16\pi G}
   \frac{Q^2}{r_+}F(\frac{1}{4},\frac{1}{2},\frac{5}{4},-\frac{Q^2}{b^2r^4_+})\equiv \frac{\omega \beta}{16
   \pi
   G}Q\Phi,
\end{equation}
where $\Phi$ is the potential at the horizon.

Third, the counter term is calculated to be
\begin{equation}
I_c=-\frac{\omega\beta_0}{16\pi
G}\int^{r_B}_{0}\Big[\frac{6r^2}{l^2}\Big]dr.
\end{equation}
Here $\beta_0$ is the time period for the  AdS space. It has to be
rescaled to match the period $\beta$ as
\begin{equation}
f(r)\beta^2 = \left(1+\frac{r^2}{l^2}\right) \beta_0^2.
\end{equation}
Then, this gives us
\begin{eqnarray}
\beta_0 &=& \beta \left[1-\frac{M l^2}{r^3}+\frac{2 b^2
l^2}{3} \left(1-\sqrt{1+\frac{Q^2}{b^2r^4}}\right)\right. \nonumber\\
&&~~~\left.\left. +\frac{2Q^2 l^2}{3r^4} {\cal
F}\left(\frac{1}{4},\frac{1}{2},\frac{5}{4}, -\frac{Q^2}{b^2
r^4}\right)-\frac{3l^4}{8r^4}\right] \right|_{r=r_B}
\end{eqnarray}
up to $r^{-4}$ after some approximations.

Finally, the background term of extremal black hole takes the form
\begin{equation}
I_e=\frac{\omega \beta M_e}{4\pi G}.
\end{equation}
Here $M_e$ is given by Eq.~(\ref{extremass}).

By carefully matching the geometries of the AdS space and the
black hole in the asymptotic region and sending $r_B$ to infinity,
the desired result for the Euclidean-Born-Infeld action can be
obtained as
\begin{eqnarray}\label{Fresult}
I_E &=& \frac{\omega\beta}{16\pi G }
   \left[r_+ -\frac{r_+^3}{l^2}- \frac{2b^2 r_+^3}{3}
   \left(1-\sqrt{1+\frac{Q^2}{b^2r_+^4 }}\right)\right.\nonumber\\
   &&~~~~~~~~~\left.+\frac{8Q^2}{3r_+}
    F(\frac{1}{4},\frac{1}{2},\frac{5}{4},-\frac{Q^2}{b^2r^4_+})-4M_e\right].
\end{eqnarray}
In deriving this expression,  we have used the condition of
$f(r_+) = 0$. As a check, we recover that  in the limit of
$b\rightarrow \infty$, the above action reduces to  the Euclidean
action for the RNAdS case~\cite{CEJM} as
\begin{equation}
I_{RNAdS} = \frac{\omega\beta}{16\pi G}
  \left(r_+ - \frac{r_+^3}{l^2}+\frac{3Q^2}{r_+}-\frac{4}{3}r_e-\frac{8}{3}\frac{Q^2}{r^2_e}
  \right).
\end{equation}
Since the free energy is given by $F=I_E/\beta$ with
$\omega=4\pi,G=1$, $I_E$ leads to  the free energy (14) in the
canonical ensemble. Hence we exactly derive the free energy which
was used  to discuss the whole phase transitions in the text.


\begin{thebibliography}{99}

\bibitem{BI} M. Born and L. Infeld,
  Proc. R. Soc. London A {\bf 144}, 425 (1934).

\bibitem{FTL} E.~S.~Fradkin and A.~A.~Tseytlin,
  Phys. Lett.  B {\bf 163}, 123 (1985).

\bibitem{Tse} A.A. Tseytlin,
  Nucl. Phys. B {\bf 276}, 391 (1986)
  [Erratum-ibid. B {\bf 291}, 876 (1987)];

\bibitem{Lei} R.~G.~Leigh,
  Mod. Phys. Lett. A {\bf 4}, 2767 (1989).

\bibitem{GSP} A. Garcia, H. Salazar and J.F. Plebanski, Nuovo. Cim {\bf 84},
 65 (1984).

\bibitem{Dem} M.~Demianski,
  Found. Phys. {\bf 16}, 187 (1986).

\bibitem{Wil} D.L. Wiltshire,
  Phys. Rev. D {\bf 38}, 2445 (1988).

\bibitem{Ras} D. A. Rasheed,
  [arXiv:hep-th/9702087].

\bibitem{Tt} T. Tamaki and T. Torii,
  Phys. Rev. D {\bf 62}, 061501 (2000).

\bibitem{GH} G. W. Gibbons and C. A. R. Herdeiro, Class. Quant. Grav. {\bf 18}, 1677 (2001).

\bibitem{Bre} N. Breton,
  Phys. Rev. D {\bf 67}, 124004 (2003).

\bibitem{AFG} M. Aiello, R. Ferraro and G. Giribet,
{\it Exact solutions of Lovelock-Born-Infeld black holes},
  Phys. Rev. D {\bf 70}, 104014 (2004).

\bibitem{Dey} T. K. Dey,
  Phys. Lett. B {\bf 595}, 484 (2004).

\bibitem{Fern}
S. Fernando,
  Phys. Rev. D {\bf 74}, 104032 (2006).

\bibitem{She}
  A.~Sheykhi,
  Phys. Lett. B {\bf 662}, 7 (2008).

\bibitem{Car} S. Carlip,
  J. Korean Phys. Soc. {\bf 28}, S447 (1995)
  [arXiv:gr-qc/9503024].

\bibitem{Man} R. Mann,
{\it Lower dimensional black holes: Inside and out},
  [arXiv:gr-qc/9501038].

\bibitem{FHR} V. Frolov, S. Hendy and A. L. Larsen,
  Nucl. Phys. B {\bf 468}, 336 (1996).

\bibitem{btz} M. Banados, C. Teitelboim and J. Zanelli,
  Phys. Rev. Lett. {\bf 69}, 1849 (1992).

\bibitem{btz1} M. Banados, M. Henneaux, C. Teitelboim and J. Zanelli,
  Phys. Rev. D {\bf 48}, 1506 (1993).

\bibitem{mtz}
  C.~Martinez, C.~Teitelboim and J.~Zanelli,
  Phys. Rev.  D {\bf 61}, 104013 (2000).

\bibitem{CG} M. Cataldo and A. Carcia,
  Phys. Lett. B {\bf 456}, 28 (1999).


\bibitem{mkp3}
  Y.~S.~Myung, Y.-W.~Kim and Y.-J.~Park,
  Phys. Rev. D {\bf 78}, 044020 (2008).

\bibitem{Ida}
  R.~Yamazaki and D.~Ida,
  Phys. Rev. D {\bf 64}, 024009 (2001).

\bibitem{HP}
  S.~W.~Hawking and D.~N.~Page,
  Commun. Math. Phys. {\bf 87}, 577 (1983).

  \bibitem{CEJM}
  A.~Chamblin, R.~Emparan, C.~V.~Johnson and R.~C.~Myers,
  Phys. Rev.  D {\bf 60}, 064018 (1999).

  \bibitem{DMMS1}
  T.~K.~Dey, S.~Mukherji, S.~Mukhopadhyay and S.~Sarkar,
  JHEP {\bf 0709}, 026 (2007).

  \bibitem{Myu}
  Y. S. Myung,
  Mod. Phys. Lett. A {\bf 23}, 667 (2008).

  \bibitem{Cho}
  Y.~M.~Cho and I.~P.~Neupane,
  Phys. Rev. D {\bf 66}, 024044 (2002).


  \bibitem{DMMS2}
  T.~K.~Dey, S.~Mukherji, S.~Mukhopadhyay and S.~Sarkar,
  JHEP {\bf 0704}, 014 (2007).

\bibitem{NO}
  S.~Nojiri and S.~D.~Odintsov,
  Phys. Lett.  B {\bf 521}, 87 (2001)
  [Erratum-ibid.\  B {\bf 542}, 301 (2002)].

\bibitem{CNO}
  M.~Cvetic, S.~Nojiri and S.~D.~Odintsov,
  Nucl. Phys.  B {\bf 628}, 295 (2002).

\bibitem{fk} S. Fernando and D. Krug,
  Gen. Relativ. Gravit. {\bf 35}, 129 (2003).

\bibitem{cpw} R.~G.~Cai, D.~W.~Pang and A.~Wang,
  Phys. Rev.  D {\bf 70}, 124034 (2004).

\bibitem{MO}
  O.~Miskovic and R.~Olea,
  Phys. Rev. D {\bf 77}, 124048 (2008).

\bibitem{GKV} D. Grumiller, W. Kummer, and D. V. Vassilevich,
Phys. Rept. {\bf 369}, 327 (2002).

\bibitem{GMc} D.~Grumiller and R.~McNees,
J. High Energy Phys. {\bf 0704}, 074 (2007).

\bibitem{MKPm}
  Y.~S.~Myung, Y.~W.~Kim and Y.~J.~Park,
  Mod. Phys. Lett. A {\bf 23}, 91 (2008).

\bibitem{abra} M. Abramowitz and I.A. Stegun,
  {\it Handbook of Mathematical Functions}, Dover, 1972.

 \bibitem{Myungplb}
  Y.~S.~Myung,
  Phys. Lett. B {\bf 638}, 515 (2006).

\bibitem{Dav} P. C. W. Davies, Proc. R. Soc. Lond. A 353, 499 (1977).

\bibitem{Pavon} D.~Pavon,
  Phys. Rev. D {\bf 43}, 2495 (1991).

  \bibitem{JP} J.~Jing and Q.~Pan,
  Phys. Lett.  B {\bf 660}, 13 (2008).

 \bibitem{Rmkp} Y.~S.~Myung, Y.-W.~Kim and Y.-J.~Park,
  Phys. Lett.  B {\bf 663}, 342 (2008).

  \bibitem{Witt} E.~Witten,
  Adv. Theor. Math. Phys. {\bf 2}, 505 (1998).

  \bibitem{GPP}
  G.~W.~Gibbons, M.~J.~Perry and C.~N.~Pope,
  Class.\ Quant.\ Grav.\  {\bf 22} 1503 (2005).


\end{thebibliography}
\end{document}